# Flexibility of short DNA helices with finite-length effect: from base pairs to tens of base pairs

Yuan-Yan Wu, Lei Bao, Xi Zhang and Zhi-Jie Tan[*]

*Department of Physics and Key Laboratory of Artificial Micro & Nano-structures of Ministry of Education, School of Physics and Technology, Wuhan University, Wuhan 430072, China*

## ABSTRACT

Flexibility of short DNA helices is important for the biological functions such as nucleosome formation and DNA-protein recognition. Recent experiments suggest that short DNAs of tens of base pairs (bps) may have apparently higher flexibility than those of kilo bps, while there is still the debate on such high flexibility. In the present work, we have studied the flexibility of short DNAs with finite-length of 5 to 50 bps by the all-atomistic molecular dynamics simulations and Monte Carlo simulations with the worm-like chain model. Our microscopic analyses reveal that short DNAs have apparently high flexibility which is attributed to the significantly strong bending and stretching flexibilities of ~6 bps at each helix end. Correspondingly, the apparent persistence length $l_p$ of short DNAs increases gradually from ~29nm to ~45nm as DNA length increases from 10 to 50 bps, in accordance with the available experimental data. Our further analyses show that the short DNAs with excluding ~6 bps at each helix end have the similar flexibility with those of kilo bps and can be described by the worm-like chain model with $l_p$~50nm.

**Keywords:** short DNA, flexibility, finite-length effect, stretching, bending, persistence length

---

[*]To whom correspondence should be addressed: zjtan@whu.edu.cn

# I. INTRODUCTION

The structural flexibility and dynamics of DNAs play important roles in their biological functions and are of fundamental importance for understanding the functions of DNAs.[1,2] In the past two decades, various advanced experimental methods have been developed to probe the structural and dynamical properties of biomolecules, and the elastic properties of long DNAs of kilo base pairs (bps) have been investigated extensively.[1-9] The existing experimental measurements and theoretical modelling indicate that a long DNA can be well described by the worm-like chain (WLC) model with a persistence length of $l_p$~50nm at physiological conditions.[1-14]

DNAs are often functional at the length of tens of base pairs,[15-23] such as the formation of nucleosome and DNA-protein recognition.[15-23] Recently, a series of experiments suggest that the flexibility of short DNAs may appear very differently from those in kilo bps.[24-28] The pioneering experiment on the cyclization of short DNAs of ~100 bps by Cloutier and Widom indicated that the cyclization occurs ~1000 times more efficiently than the prediction from the WLC model.[24] The atomic force microscopy (AFM) experiment by Wiggins et al showed that the bending of DNAs at short length scale is larger than that described by the WLC model for DNAs at long length scale.[25] Another experiment with fluorescence resonance energy transfer (FRET) and small angle x-ray scattering (SAXS) by Yuan et al also suggested the higher flexibility of short DNAs of 15-89 bps which is beyond the description of the conventional WLC model.[26] The recent SAXS experiment on short DNAs of ≤35 bps with two linked gold nanocrystals by Mathew-Fenn et al has suggested that short DNAs are at least one order of magnitude more extensible than long DNAs of kilo bps revealed by the previous single molecule stretching experiments.[27] The very recent experiment on the cyclization of short DNAs of 67-106 bps by Vafabakhsh and Ha showed that the looping rate of short DNAs could not be well described by the WLC model.[28] To understand such high flexibility of short DNAs, some possible mechanisms have been proposed based on the experimental findings.[25-27] Yuan et al modified the WLC model by incorporating base-pair-level fluctuation and found that the possible base-pair flip out could lead to the high flexibility of short DNAs.[26] Wiggins et al suggested that spontaneous large-angle bends may be responsible for the high flexibility at short length scale.[25] Mathew-Fenn et al proposed a mechanism of the long-ranged stretching cooperation over two turns.[27]

However, another series of experiments and analyses show the opposite conclusion. Mastroianni et al have studied the flexibility of short DNAs with finite-length of 42-92 bps with two linked gold nanocrystals by SAXS and the WLC model, and concluded that the short DNAs of 42-92 bps can be described by the WLC model with persistence length of ~50nm.[29] Recently, Vologodskii and Frank-Kamenetskii[30,31] analyzed the experiment by Cloutier and Widom[24] and that by Vafabakhsh and Ha[28]. Their analyses showed that, for the former experiment,[24] the very high ligase concentration used in the experiment was considered to be responsible for the discrepancy with the WLC model,[30,31] and for the latter experiment,[28] the possible sequence error in synthetic oligos and long sticky ends very likely resulted in the significant difference from the WLC model[30,31]. Very recently, Mazur and Maaloum[32,33] employed the AFM-in-solution with high accuracy to re-examine the experiment of Wiggins et al[25], and showed that the flexibility of DNAs at length scale ≳ two turns can be well described by the WLC model.[32,33]

Up to now, the previous experiments and analyses might have already led to the conclusion that the flexibility of a short DNA at length scale approximately longer than two helical turn can be described by the WLC model with persistence length of ~50nm.[2-5,13,32,33] However, for the global flexibility of short DNAs with tens of bps,[24-31] there is still lack of consensus. First, the recent analyses did not agree with the extremely high flexibility of short DNAs with length from ~60 to ~100 bps observed in the cyclization experiments.[24,28,30,31] Second, the SAXS/FRET experiments showed apparently higher flexibility for short DNAs with length ≤35 bps,[27] while other SAXS experiments indicated that the flexibility of short DNAs of 42-92 bps can be described with the WLC with persistence length of ~50nm.[29] Therefore, there is still the controversy on the global flexibility of short DNAs with tens of bps.[24-31] To understand the elusive controversy and the relevant mechanism, we will focus on the flexibility of short DNAs with length ≤50 bps.

Due to the small size of short DNA, it is rather difficult to accurately characterize their structure flexibility by experiments, since certain inaccuracy would be generally involved via various experimental techniques[24-33] and may become nonnegligible for small-sized DNAs. For example, the SAXS methods generally involve two linked gold nanocrystals which can be comparable to short DNAs in size,[26,27,29] and the AFM methods generally involve DNAs of kilo bps attached on a substrate which would generally ignore the finite-length effect in DNA flexibility.[25,32,33] To circumvent the difficulty in studying short DNAs, the all-atomistic molecular dynamics (MD) simulation could be a useful tool to probe and analyze the flexibility of short DNAs and RNAs.[35-40] Recently, Noy and Golestanian have employed 130ns MD simulations with SPC/E water model to investigate length-scale dependence of DNA mechanical properties, where they adopted two DNAs of 56 and 36 bps.[35] However, they did not obtain length-dependent flexibility of short DNAs since the segments at helix ends were ignored in the data analysis and only two short DNAs were used in the study.[35-37] Therefore, it is still necessarily required to study the flexibility of short DNAs with tens of base pairs.[24-31]

To study the length-dependent flexibility of short DNAs, we employ the all-atomistic molecular dynamics (MD) simulations and Monte Carlo (MC) simulations with the WLC model, to systematically investigate the flexibility of 8 short DNAs of 5 to 50 bps in 1M NaCl solutions. The finite-length effect will be taken into account in the present work, since such effect was naturally included in the relevant experiments[24,26-29] and was generally ignored in previous MD and AFM studies.[32,33-37] Firstly, we will analyze the global flexibility of short DNAs with various lengths through calculating the distributions of contour length and end-to-end distance. Afterwards, we will analyze the detailed structure flexibility at base-pair level by calculating helical rise and bending angle, helical twist for short DNAs of different lengths. Additionally, we will calculate the apparent persistence length for various short DNAs and make comparisons with the existing studies. Finally, we will quantitatively discuss the finite-length effect in the flexibility of short DNAs of different lengths. Beyond the recent studies,[35-38] the present work will be focused on the flexibility of short DNAs of finite length rather than the local flexibility of a DNA at short length-scale, and will include 8 short DNAs with the wide length range from 5 to 50 bps to obtain the length-dependent flexibility of short DNAs.

## II. MODEL AND METHOD

### A. All-atomistic molecular dynamics and data analysis

The short DNAs of finite-length used in the study are of 5, 10, 15, 20, 25, 30, 40 and 50 bps which are displayed in Fig. S1 of supplementary material.[41] The sequences of the short DNAs are selected according to the sequences in recent experiments to yield normal B-form DNA helices,[27] and contain all the dinucleotide base pairs[42,43]; see Table 1. All the DNA strands are perfectly complementary.

The initial structures of short DNA helices were taken as the standard B-DNA fibers which were immersed in rectangle boxes containing explicit water and ions. The rectangle boxes for the short DNAs of different lengths are different and listed in Table 1, as well as the melting temperature calculated from the nearest neighbor model.[44-46] It is noted that the melting temperatures of the short DNAs are significantly higher than room temperature (298K) and consequently the short DNAs would generally keep their rather stable helix structures. The counterions of $Na^+$ and 1M NaCl salt ions were added with Amber LEaP Program[47,48] to ensure that the negatively charged backbone of DNA is nearly full-neutralized.[45,46] The efficient Particle-Mesh-Ewald method was employed for treating the electrostatic interactions, and the cut-off length of 1.4nm was used for treating long-range interactions.[49] The periodic boundary condition was also employed in all the MD simulations.[49] In each MD simulation, a DNA was initially placed in the center of simulation cell and its axis was initially kept in parallel with z-axis. The distance between DNA and box edges was initially kept larger than at least 1nm in z-axis. In the simulations, we used the Amber parmbsc0 force field and the TIP3P water model combined with parmbsc0 ion model,[50-52] since the force field has been shown to give good description for various nucleic acids[35-40,53,54] and the usage of TIP3P water model has been previously shown to be more efficient in convergence than the SPC/E water model[35-37].

All systems were optimized, thermalized and equilibrated with Gromacs 4.5[49,55] as follows. Firstly, an energy minimization of 5,000 steps was performed with the steepest descent algorithm at low temperature, and then the systems were slowly heated to 298K and equilibrated with the Nose-Hoover temperature coupling until 0.5 ns. Afterwards, the subsequent NPT simulations of 20 ns (time-step 1fs, P=1atm) were performed with the Parrinello-Rahman pressure coupling and with the short DNAs released. Finally, the simulations of the systems were continued for another 200 ns in the isothermic-isobaric ensemble (P=1atm, T=298K).[56] A time step of 1fs was used in the conjunction with a Leap-frog algorithm,[57] for capturing the detailed dynamic motion of atoms. For each short DNA, three independent all-atomic MD simulations were performed. To test the convergence of the MD simulations, we have examined the instantaneous value of end-to-end distance versus MD time for two independent simulations until 300ns for the 50-bp DNA helix, as shown in Fig. 1 & Fig. S2 in supplementary material.[41] We found that the MD simulations are nearly converged after ~100ns for the 50-bp DNA, since the relative difference between the mean values of end-to-end distance in the time ranges of [100ns-200ns] and [100ns-300ns] is only ~0.1%. We have examined the MD trajectories and found that the short DNAs (e.g. 50-bp DNA) mainly exhibit the conformational fluctuations with apparent bending and moderate rotation rather than sharply rotate to the direction of short box boundary; see the movies in supplementary material.[41] Furthermore, for all the MD trajectories used here, we have calculated the minimum distances between the two end segments of short DNAs and found that they are always larger than the cut-off length of 1.4nm, as shown in Fig. S2 in supplementary material.[41] In addition, we have examined the effect of simulation box size by performing an additional simulation for a larger box, and the box-size effect appears negligible, as shown in supplementary material.[41]

In analyzing the conformational change of short DNAs at base-pair level, we first located a local coordinate system with an orthonormal basis.[58] Afterwards, the central axes of DNA helices were derived with the use of the program Curves+,[59] and then were used for analyzing the length-dependent structural properties of short DNAs, such as contour length, end-to-end distance, helical rise, helical radius, helical twist, bending and persistence length. In the simulations, due to the very high melting temperature of the used short DNAs (see Table 1), we did not observe the opening of the inner base pairs with all (two for A·T and three for C·G[1]) hydrogen bond (H-bond) opening while the terminal base pair at each helix end could occasionally become open. Such few conformations with the opening of two terminal base pairs have been ignored in the data analysis with the program Curves+.[59]

## B. Monte Carlo with worm-like chain model

In addition to the above described all-atomistic MD simulations, we have performed the MC simulations with the WLC model. In the model, a $N$-bp short DNA helix is modeled as a simplified linear chain of $N$ sequential beads, and each bead represents one base pair. The energy $U$ of the chain is composed of two contributions: the bond-angle energy and bond-length energy which describe the bending and stretching rigidities, respectively. Following the previous studies,[29,60] the bond-angle energy at a bead can be given by[29,60,61]

$$U_b = k_B T \frac{l_p}{l_0}(1-\cos\theta),\qquad(1)$$

where $l_p$ is the persistence length of the chain. $l_0$ is the mean bond length between two adjacent beads, and $\theta$ is the local bending angle between the adjacent unit vectors. Our calculations showed that the form of Eq. 1 gives the similar results with a harmonic form[62] (data not shown). The bond-length energy for a certain bond length $l$ can be expressed as[29,60-62]

$$U_s = \frac{1}{2}\frac{k}{l_0}(l-l_0)^2,\qquad(2)$$

where $k$ is the stretching modulus of the chain.

Based on the above bond-angle and bond-length energies, the MC method was employed to sample the WLC model of short DNAs with pivot move algorithm which has been shown to be rather efficient in sampling the conformations of a polymer.[63-65] The conformation sampling of the bead chain was performed according to the standard Metropolis algorithm.[64,65] In each MC simulation, the WLC model could get equilibrated within $10^4$ steps, and the following $2\times 10^7$ steps were employed to calculate the averaged properties of the WLC model. The efficient pivot move and enough equilibrium MC steps would ensure the calculations of equilibrium properties for the WLC model.

## III. RESULTS AND DISCUSSION

In this section, we will analyze the flexibility of short DNAs with finite-length effect based on the all-atomistic MD and MC simulations with the WLC model, and 8 short DNAs of 5-50 bps will be covered. Firstly, we will focus on the global conformation fluctuation of short DNAs and afterwards, we will make detailed analysis on length-dependent properties of helical structure at base-pair level. Additionally, we will calculate the apparent persistence length of various short DNAs, and make extensive comparisons with the previous studies. Finally, we will quantitatively discuss the finite-length effect in DNA flexibility.

### A. Global structural flexibility

The global size of a short DNA can be characterized by its contour length $L$ and end-to-end distance $R_{ee}$. As shown in Fig. 1, to include the finite-length effect, the contour length $L$ is taken as the summation over all base pair steps and the end-to-end distance $R_{ee}$ is taken as the distance between the centers of two terminal base pairs. The variance $\sigma_L^2$ of contour length of a short DNA is expressed by $\sigma_L^2 = \overline{(L-\bar{L})^2}$, and the variance $\sigma_R^2$ of $R_{ee}$ is given by $\sigma_R^2 = \overline{(R_{ee}-\overline{R_{ee}})^2}$.

As shown in Fig. 2a, the MD trajectories for short DNAs of various lengths give the distributions of contour length $L$ and end-to-end distance $R_{ee}$ with approximately symmetric fluctuations around their mean distance. It is noted here that the distributions of $R_{ee}$ slightly deviate from the normal distribution for the 40-bp and 50-bp DNAs, which is in accordance with the previous WLC analysis for semiflexible polymers with $L<l_p$.[66,67] We have examined the possible statistical error and found that the relative difference between the maximum value and mean value of $R_{ee}$ is small (<1% for the 50-bp DNA). Due to the small deviation and the wide usage in previous related analysis,[26,27,29] we still used the mean value and its variance to make analysis for the flexibility of the 40-bp and 50-bp DNAs. As shown in Fig. 2b, the mean contour length $\bar{L}$ and mean end-to-end distance $\overline{R_{ee}}$ increase approximately linearly with DNA length. The slope for $\bar{L}$ with DNA length gives an average rise of ~3.32Å per base pair step, in agreement with the crystallographic value of 3.32±0.19Å,[68] and the slope of $\overline{R_{ee}}$ with DNA length gives an average value of ~3.22Å per base pair which is also close to the value of 3.27±0.1Å in the recent SAXS experiment[27]. The structural fluctuations of a short DNA can be quantified by the variances $\sigma_L^2$ for $L$ and $\sigma_R^2$ for $R_{ee}$, respectively. Fig. 3 shows the variances $\sigma_L^2$ and $\sigma_R^2$ as functions of DNA length, which were derived from the all-atomistic MD simulations. As shown in Figs. 3a and b, $\sigma_L^2$ increases roughly linearly with DNA length $N$ over the range of 5-50 bps. Fig. 3c shows that $\sigma_R^2$, the variance of $R_{ee}$ increases quadratically with DNA length, which is qualitatively in accordance with the recent SAXS experiments with two attached gold nanocrystal.[27] Since the force field associated with the thioglucose-passivated gold nanocrystals involved in the experiments is unavailable,[69-71] the strict and direct comparisons with the experimental data[27] are absent in the present study and a non-strict comparison with the experiments is given in Figs. S3 and S4 of supplementary material.[41] Nevertheless, the accurate treatment on short DNAs with linked nanocrystals and the direct comparison are still required in future works, not only due to examining the modeling results, but also due to understanding the effect of linked nanocrystals on the flexibility of short DNAs.

The Monte Carlo simulations with the WLC model have also been performed for short DNAs, in a comparison with the results from the all-atomistic MD simulations. As shown in Eqs. 1 and 2, there are two (bending and stretching) parameters for the WLC model: persistence length $l_p$ and stretching modulus $k$ which describe the global flexibility of a DNA. The previous combination of experiments and the WLC model gives $l_p$~50nm and $k$~1400-1600pN for long DNAs of kilo bps at high salt solutions.[1-7,12,72,73] As shown in Figs. 3a and b, the variance $\sigma_L^2$ of contour length $L$ is not sensitive to persistence length $l_p$ while sensitive to stretching modulus $k$. This is reasonable since $l_p$ mainly reflects bending rigidity while $k$ describes stretching rigidity, and contour length $L$ mainly depends on the latter. As shown in Figs. 3b and c, the WLC model with the parameters ($l_p$~50nm and $k$~1500pN) for long DNAs apparently underestimates the conformational fluctuation of short DNAs. Figure 3b also shows that, over the range of 10–50 bps, the length-dependent $\sigma_L^2$ curve of WLC model is closest to that of all-atomistic MD at $k$~1320pN, which is not far away from the experimental value of 1400-1600pN for long DNAs in high salt solutions.[3,7,72] By taking $k$~1320pN and different $l_p$'s, we found that the short DNAs of ≤50 bps have higher bending flexibility than long ones of kilo bps, which is reflected by the variance $\sigma_R^2$ of end-to-end distance $R_{ee}$ in Fig. 3c; see also Fig. S5 for the distribution of $R_{ee}$.[41] In addition, Fig. 3c shows that for $l_p$ = 45nm (and $k$=1320pN), $\sigma_R^2$ of the WLC model for the 50-bp DNA appears close to that of the all-atomistic MD. For the DNAs with length <50 bps, the WLC model with $l_p$=45nm and $k$=1320pN still slightly underestimates the conformation fluctuation, indicating higher flexibility for shorter DNAs. For example, as shown in the inset of Fig. 3c, the WLC model with $l_p$ = 30nm (and $k$= 1320pN) still visibly underestimates $\sigma_R^2$ for DNAs length ≲10 bps as compared with the all-atomistic MD simulations. It is noted that the WLC model does indeed reproduce the desired $l_p$ with a high degree of accuracy, as shown in supplementary material.[41]

The above analysis on the global conformation fluctuation of short DNAs of different lengths shows that shorter DNAs have higher global flexibility (lower apparent persistence length) and with a certain pair of persistence length $l_p$ and stretching modulus $k$, the WLC model cannot reproduce the results of all-atomistic MD over the DNA length range from 5 to 50 bps. Since stretching and bending are two major contributions to the global flexibility of short DNA, in the following, we will analyze the helical stretching and bending of short DNAs at base-pair level.

**B. Helix stretching**

The helix stretching of short DNAs may be characterized by three parameters: rise per bp, helix radius and twist angle per bp.[58,59] In the following, we will analyze the length-dependence of rise, helix radius and twist angle based on the MD trajectories for short DNAs of different lengths. Our analysis shows that helix rise and radius are strongly correlated, i.e., the increase and decrease of rise are generally accompanied with the decrease and increase of helical radius, respectively; see Fig. S6a of supplementary material.[41] That is to say, the helix stretching/shrinking of a DNA is generally accompanied by helix thinning/thickening. It is reasonable since helix stretching/shrinking would lead to the less/more crowding of atoms in a DNA, and consequently causes DNA helix thinning/thickening.

Firstly, we calculated the distribution of base-pair rise and rise variance along the 20-bp, 30-bp, and 50-bp DNAs, respectively. As shown in Figs. 4a and b, in spite of the fluctuation of base-pair rise between ~3Å and ~3.6Å, the rises of base pairs near two helix ends appear slightly larger than those in the middle of short DNAs. Moreover, the variance of rise also has slightly larger values for the base pairs near two DNA ends than that for the middle base pairs. The larger variance of rise near the two ends suggests that the segments at two ends have stronger stretching flexibility than those middle base pairs, which would cause the length-dependent stretching flexibility for short DNAs. As shown in Figs. 4c and d, with the decrease of DNA length, mean rise and rise variance increase slightly, suggesting slightly higher stretching flexibility for shorter DNAs. The mean rise per bp over different short DNAs of 10-50 bps is close to the value of 3.32±0.19Å from the crystallographic experiments.[68,74]

Corresponding to the negative correlation between helix rise and radius, the segments near two helix ends have smaller helical radius and larger variance, as shown in Fig. S8a of supplementary material.[41] Consequently, the mean helix radius increases slightly from ~9.3Å to ~9.6Å and its variance decreases, as DNA length increases from 10 to 50 bps. Therefore, the analyses on helix rise and radius show that shorter DNAs have stronger stretching-shrinking conformation fluctuation and the mean values of helix rise and radius are ~3.32Å and ~9.5Å over 10-50 bps which are close to those of the experiments.[68-75] It is understandable that shorter DNA has higher stretching flexibility because of the less spatial constraints for helix stretching/shrinking and consequently higher stretching flexibility for the segments at two helix ends.

In addition to helix rise and radius, we also examined helix twisting. Our analysis shows that there is strong fluctuation of twist angle, and the mean twist angle per base pair increases very slightly from ~32º to ~33º when DNA length increases from 10 to 50 bps; see Fig. S9 of supplementary material.[41] Additionally, the variance of twist angle also appears larger for the base pairs near two helix ends in spite of fluctuation. It is reasonable since the segments at helix ends have less spatial constraints to retain the helical structure. Furthermore, we examined the coupling between stretching and twisting. Our analysis shows that helix rise and twist are strongly coupled, i.e., the increase and decrease of rise are generally accompanied with the increase and decrease of twist angle, respectively; see Fig. S6b in supplementary material.[41] Such negative twist-stretch coupling is in accordance with the previous studies.[72,76,77]

In the subsection of "Global structural flexibility", we mainly used the helical axis of a short

DNA to characterize contour length $L$ and end-to-end distance $R_{ee}$ which quantify the global flexibility of a short DNA as a linear polymer. The length-dependence of helix twisting shows that the slightly smaller twist angle and larger variance of twist angle may also contribute to the global flexibility of shorter DNAs.[76,77]

**C. Helix bending and sharp bending**

*1. Helix bending*

In addition to helix stretching, helix bending is another major contribution to the structure flexibility of DNAs.[30,31,78-83] Since bending is difficult to be characterized at base-pair level, we use the bending over 6 base pairs to characterize the helical bending of short DNAs,[79,80] and the bending angle over 6 bps is calculated as the angle between the first unit vector and the last unit vector along a DNA axis over adjacent 6 base pairs.[79] At first, we calculated the distributions of bending angle (over 6 bps) and bending angle variance along short DNAs of 20, 30, and 50 bps. As shown in Figs. 5a and b, the bending angle and its variance near two helix ends are distinctly larger than those in the middle of short DNAs, suggesting that the base pairs near two ends have significantly stronger bending flexibility. Beyond the bending distribution along short DNAs, we calculated the mean bending angle and its variance as functions of DNA length. As shown in Figs. 5c and d, corresponding to the stronger bending flexibility near two ends described above, the mean bending angle and its variance both increase apparently as DNA length decreases, suggesting the stronger bending flexibility for shorter DNAs. Physically, the segments near two helix ends of short DNAs have stronger bending flexibility due to less (more) spatial constraints (freedom) for bending, which would cause the stronger bending flexibility for shorter DNAs.

*2. Helix sharp bending*

To further understand the bending properties of a short DNA, we analyzed the degree of helix bending along DNAs. Here, a kind of sharp bending was defined as those bending angles with >30º over 6 base pairs. Figure 6 shows that, along a short DNA, the segments near the helix ends are more likely to sharply bend and the mean probability of sharp bending (per 6 bps) for a short DNA decreases with the increase of DNA length. Physically, due to the less spatial constraints from outer segments at two ends, the segments near DNA ends would fluctuate more strongly and frequently, and sharp bending would occur more probably near two ends. As a result, the bending angle is larger for shorter DNAs. In addition, we calculated the probability of such sharp bending for the WLC model with the same apparent persistence length. As shown in Fig. 6a, the probability of sharp bending from the all-atomistic MD is apparently higher than that of the WLC near two helix ends, suggesting that the bending properties along a short DNA are beyond the description of the WLC model.

What causes the sharp bending of a free DNA? Is sharp bending caused by the opening of H-bonds? To classify the microscopic mechanism, we analyzed the status of H-bonds between base-pairing nucleotides when a finite-length DNA sharply bends. Firstly, we examined the averaged H-bond opening probability along the 40-bp DNA, where we used the criteria of bond length >3.5Å to determine the H-bond opening.[84] As shown in Fig. S10 of supplementary material,[41] only H-bonds at major groove have visible opening probability, while those at minor groove and between two grooves (for C·G base pairs) have very low (almost invisible) opening probability. The two terminal base pairs have higher H-bond (at major groove) opening probability than those in the middle helix, and A·T base pairs have higher H-bond opening probability than C·G base pairs. Physically, the terminal base pairs have no constraints from outer base pairs and thus could fluctuate more strongly. Consequently, the two terminal base pairs have higher opening probability of H-bond. In addition, C·G base pairs have one more H-bond than A·T, thus the H-bond of C·G base pairs has more spatial constraints and consequently has low opening probability than that of A·T base pairs. However, there is no visible higher H-bond opening probability near the sharp bending sites, and for a sharp bent conformation, H-bonds are not disrupted at the sharp bending sites, which suggests that H-bond opening may not make the major contribution to the sharp bending for short DNAs. It is reasonable since the melting temperatures of the short DNAs are significantly higher than room temperature (298K); see Table 1.

In another way, the microscopic analysis on rise distribution along helix shows that at the junction between stretching and shrinking segments on a DNA, the helix is shown to sharply bend. As shown in Figs. 6c and d, there are two stretching-shrinking junctions along the 40-bp DNA, and there are two sharp bending sites at the two corresponding junctions. Furthermore, our microscopic analysis shows that, sharp bending is generally towards major groove and during bending, the base pairs at bending sites would slide away from helix central axis to minor groove and major groove would become deeper, which favors a sharp bending toward major groove in addition to the broad width of major groove.[79,80] Furthermore, such sharp bending would be aided by the asymmetrical ion binding to bent DNAs.

*3. Asymmetrical ion binding aids bending*

Since DNAs are negatively charged macromolecules, metal ions can play an important role in DNA structure deformation such as helix bending.[79,80,85-96] As shown in Fig. 7, $Na^+$ ions become condensed around negatively charged DNA helix, and at the base pairs where the helix is sharply bent, there are more $Na^+$ ions accumulated to screen the Coulombic repulsion due to the bending. To further analyze the role of $Na^+$ during helix bending, we separated the condensed $Na^+$ ions into those binding in bending direction and in the opposite direction. As shown in Figs. 7c and d, it is clear that the local density of condensed $Na^+$ ions in bending direction is higher than that in the opposite direction. Such asymmetrical ion binding would assist DNA bending, which is also in accordance with the previous analysis.[79,80,95] The salt effect, including ion concentration and ion valence, on DNA bending is very important and deserved to be investigated separately in the future.

**D. Apparent persistence length of short DNAs**

Like other polymers, the global flexibility of a DNA can be quantified by its persistence length $l_p$.[2-7,12,13,97-103] In experiments, the persistence length of a DNA can be determined by fitting an elasticity model to stretching experimental data,[2-5,12] or by measuring the bending angle along the tangent vectors of a DNA[25,32,33,99] or by measuring the size of a DNA.[26,27,98,100] In the following, we will calculate apparent persistence length of short DNAs with different lengths since DNA ends have higher flexibility than the central base pairs. Based on the MD trajectories and the program Curves+,[59] the apparent persistence length $l_p$ for a short DNA of $N$ bps can be calculated by [65,101,102]

$$l_p = -L(\ln(\langle \hat{u}(1) \cdot \hat{u}(N-1) \rangle))^{-1}, \qquad (3)$$

where $\hat{u}$ is the unit vector along the central axis of a DNA which connects the centers of adjacent base pairs. Meanwhile, the apparent persistence length for a short DNA can also be calculated through the statistics of the mean square end-to-end distance $\langle R_{ee}^2 \rangle$ by[2,65]

$$\langle R_{ee}^2 \rangle = 2l_p L \left(1 - l_p(1 - \exp(-L/l_p))/L\right). \qquad (4)$$

The two equations (Eqs. 3 and 4) were both employed in calculating $l_p$ for different short DNAs. Figure 8a shows $l_p$ as a function of DNA length. With the increase of DNA length from 10 to 50 bps, $l_p$ increases gradually from ~29nm to ~45nm and the two methods (Eqs. 3 and 4) give the similar predictions on $l_p$. For longer DNAs, $l_p$ is expected to become close to the value (~50nm) of long DNAs of kilo bps.[2-5,13,34] Our calculations also indicate that, $l_p$'s calculated from MC with the WLC model are just close to that set in the model (Eq. 1) and nearly independent of DNA length. This suggests that the WLC model with a certain pair of parameters (persistence length $l_p$ and stretching modulus $k$ in Eqs. 1 and 2) cannot describe the length-dependent apparent persistence length for short DNAs of different lengths.

*1. Comparisons with previous studies*

Firstly, we make comparisons with experiment data.[26,29,35,80] Yuan et al have employed FRET and SAXS to quantify the radius of gyration $R_g$ of short DNAs with 16, 21, and 66 bps, and $R_g$ can be used to estimate the apparent persistence length[26]; see the caption of Fig. 8. As shown in Fig. 8, the predicted $l_p$'s agree well with the data from the experiments, except for $N$=16 bps. Our value of ~32nm for $N$=15 is higher than that (~16nm) of the experiments for $N$=16.[26] Such deviation for the 16-bp DNA may possibly arise from the fact that the inaccuracy involved in the experiments may become stronger for shorter DNA and the calculation of $l_p$ is sensitive to the measurement of $R_g$. Also, the difference in sequence may contribute to the deviation. Mastroianni et al have also employed SAXS to probe the conformation flexibility of DNAs of 42-94 bps with two linked end nanocrystals.[29] By combining with the WLC model, they concluded that the studied short DNAs have approximately $l_p$ of ~50nm. Our prediction on $l_p$ does not differ much from the experimental value since our predicted $l_p \gtrsim$43nm for $N \geq$40 bps. Most of AFM experiments were performed for long DNAs in kilo bps attached on a substrate and thus, the segments at two helix ends have generally been ignored in the AFM measurements.[25,32,33,103] Consequently, a direct comparison on $l_p$ with the AFM experiments is absent in the present work. More extensive comparisons with experiments require more accurate experiment data on $l_p$ for short DNAs. Secondly, we make the comparisons with the previous simulational studies.[35,80] Noy and Golestanian's all-atomistic MD gives that, the averaged apparent persistence length is ~43nm for a 56-bp DNA,[35] a value close to our prediction (~45nm for 50-bp). Spiriti *et al* employed the adaptive umbrella method to simulate DNA bending and found that the 12-bp DNAs with 33% A·T bps and excluding two terminal bps have a $l_p$ of ~41nm at 150mM KCl.[80] This value is a little higher than our value of ~32nm for the 15-bp with 40% A·T bps at 1M NaCl. Such slightly higher value of $l_p$ of Spiriti et al may come from the exclusion of two terminal base pairs in the data analysis and much lower bulk salt employed in the simulations.[80]

*2. Empirical formula*

Based on the apparent persistence lengths from the all-atomistic MD for the short DNAs of 10-50 bps in 1M NaCl solutions, we obtained an empirical formula for $l_p$ as a function of DNA length $N$ (bps)

$$l_p(N) = l_p^0 - A/(B+N), \tag{5}$$

where $l_p^0$ (~50nm) is the persistence length of DNAs of kilo bps, and $N$ is DNA length (bps). The parameters $A$~450nm and $B$~10. As shown in Fig. 8, Eq. 5 fits the values from the all-atomistic MD and experimental data well and would approach gradually to the value of long DNA as DNA length increases. Here, it would be interesting to revisit the variance $\sigma_R^2$ of $R_{ee}$ discussed in the subsection of "Global structure flexibility". Can we reproduce the length-dependent $\sigma_R^2$ of all-atomistic MD by the WLC model with the length-dependent apparent persistence length $l_p$? As shown in Fig. 8b, with the apparent persistence length $l_p$ from Eq. 5 and stretching modulus $k$=1320pN, $\sigma_R^2$'s from the WLC model are close to those of the all-atomistic MD over the DNA length range from 10 to 50 bps.

**E. Finite-length effect**

In the above, we have shown the apparently stronger bending and stretching flexibilities near two ends of short DNAs, which causes the distinct finite-length effect in DNA flexibility. To quantitatively examine the finite-length effect in the flexibility of short DNA, we make another analysis on the all-atomistic MD data by comparing the variance $\sigma_R^2$ of isolated short DNAs and that of the "inner" short DNAs with the same length which are taken from the middle of the 50-bp DNA. As shown in Fig. 9a, we found that the isolated short DNAs have apparently higher structural flexibility ($\sigma_R^2$) than those "inner" short DNAs with the same length, verifying the higher flexibility of shorter DNA which is attributed to the significantly higher flexibility of the segments at helix ends.

Then how many base pairs at each helix end contribute to the higher flexibility of short DNAs? In the following, we will quantify the number of base pairs which contribute to the strong finite-length effect. Firstly, we calculated the variance $\sigma_R^2$ of end-to-end distance for the short DNAs with excluding several (2, 4, 6, 8 and 10) base pairs at each end, and compared $\sigma_R^2$ with the WLC model with $l_p$=50nm (and $k$=1320pN). As shown in Fig. 9b, with the increase of the number of excluded end base pairs, $\sigma_R^2$ of short DNAs decreases and approaches to that of the WLC model, and when ~6 base pairs at each end are excluded from DNAs, $\sigma_R^2$ appears almost identical to that of the WLC model with $l_p$=50nm. For these short DNAs with excluding end base pairs, we also calculated apparent persistence length $l_p$ through Eq. 3. As shown in Fig. 9c, as more end base pairs are excluded, $l_p$ would increase and when more than ~6 end base pairs are excluded, $l_p$ for short DNAs of different lengths would converge to ~50nm, the value of long DNAs in kilo bps.

Therefore, Fig. 9 suggests that, the ~6 base pairs at each helix end are responsible for the finite-length effect which results in the significantly higher flexibility (lower apparent $l_p$) of short DNAs. This is reasonable since base pairs at helix ends generally fluctuate with less spatial constraints and consequently have the large bending/stretching flexibility. We expect that the finite-length effect would become weak when DNA length is much longer than 6 bps, and the number of ~6 (bps) would change when the solution ionic condition deviates away from 1M NaCl, due to the significant electrostatic contribution to DNA flexibility.[13,79,85-98,104]

## IV. CONCLUSIONS

In this paper, we have studied the flexibility of short DNAs with finite-length effect by the all-atomistic MD simulations and MC simulations with the WLC model. The length of short DNAs employed in the study covered the wide range from 5 to 50 bps. We investigated the flexibility of short DNAs of various lengths by analyzing the end-to-end distance and contour length, and by analyzing the stretching and bending of short DNAs of different lengths at base-pair level. The main conclusions are as follows: (i) The short DNAs have lower apparent persistence length than long ones, and such low apparent persistence length is attributed to the high flexibility of ~6 base pairs at each helix end; (ii) Due to the strong finite-length effect caused by the high flexibility near two helix ends, the WLC model with the persistence length (~50nm) of long DNAs in kilo bps would underestimate the structure flexibility of short DNAs of 5-50 bps, and only with excluding ~6 base pairs at each end, a short DNA can be described by the WLC model with persistence length of ~50nm; (iii) The short DNAs may sharply bend at stretching-shrinking junctions, and such sharp bending occurs more frequently near two helix ends; (iv) The apparent persistence length of short DNAs with finite-length effect increases gradually from ~29nm to ~45nm as DNA length increases from 10 to 50 bps, and we obtained an empirical formula for the apparent persistence length as a function of DNA length which may be practically useful.

In addition to the above described general conclusions, the present work would be very helpful for understanding the elusive controversy on the global flexibility of short DNAs arising from the existing experiments.[24-31] Our results show that, compared with long DNAs, short DNAs have apparently higher flexibility for DNA length $N \leq$ ~30 bps while slightly higher flexibility for $N \geq$ ~40 bps, which might help to bridge the gap between the observations from the experiments of Mathew-Fenn et al for $N \leq 35$ bps[27] and those of Mastroianni et al for $N > 40$ bps.[29] The nanocrystals linked with short DNAs in the experiments may also contribute to the experimental observations,[26,27,29,105,106] while was not involved in the present work. However, the present work could not explain the cyclization experiments of short DNAs with ~60-100 bps[24,28] since the end effect would become small for DNAs of ~60-100 bps. Furthermore, DNA cyclization generally involves the formation of DNA loop in which the ends are not fully free.[24,28,30,31]

The present work is apparently different from the previous MD works, e.g., the work of Noy and Golestanian,[35] at least in the following four aspects: (i) The present work covers the wide length range of short DNAs to systematically study the length-dependent global flexibility of short DNA with finite-length effect which was often ignored in the works of Noy and Golestanian and others [25,32,33,35]; (ii) The length-dependent apparent persistence length of short DNAs has been obtained in the work; (iii) The WLC model has been extensively employed in parallel with the all-atomistic MDs, to make more comprehensive analyses on the flexibility of short DNAs; (iv) The present work is extended to longer simulation time and three independent MD runs to warrant the converge and the stable equilibrium properties. Moreover, the conclusion that shorter DNAs have higher flexibility (lower apparent persistence length) is not contradictory with the recent AFM experiments (e.g. that of Mazur and Maaloum[32,33]) since it is the segments near helix ends which results in the higher flexibility of short DNAs while such effect was generally ignored in the AFM studies.[25,32,33,103] Our analyses show that the flexibility of short DNAs with excluding ~6 base pairs at each end would behave like long DNAs of kilo bps, and the finite-length effect would become weak for long DNAs.

The present work also has some approximations and limitations. Firstly, we employed the quadratic function in analyzing the distributions of bending and its variance (and rise and its variance) along short DNAs in spite of certain fluctuation. However, the overall larger values of bending angle and variances of bending angle and rise at helix ends for all short DNAs would indeed suggest the higher flexibility of DNA ends. Secondly, the analyses and predictions in the work would strongly depend on the force field employed in the MD simulations, while our analyses show that the employed force field should be reliable since the predicted apparent persistence length approaches to ~50nm (the well-accepted experimental value) for the short DNAs without >~6 base pairs at each helix end. Thirdly, we ignored the few conformations with the terminal base pair opening since the program Curves+ would produce abnormal analysis for such conformations, while such ignorance should not notably affect our results due to the very small amount of the conformations. The inclusion of such states with frayed terminal base pairs at the ends of short DNAs would further increase the flexibility of base pairs at two ends. Fourthly, although there is the direct comparison with experimental persistence length, the present work did not involve the direct comparisons with the experiment on length-dependent end-to-end distance and its variance due to the lack of the associated force field for the linked nanocrystals[27]. Such direct and strict comparison is still necessary and significant in the future study, which would be helpful for examining the simulation analysis, and understanding the effect of the labeling nanocrystals on probing the flexibility of biomacromolecules.[105,106] Finally, the flexibility of DNAs is also dependent on sequence,[41,42,45,107] temperature[108,109] and ionic condition[5,13,92,99], which is beyond the scope of the present work and would be deserved to be discussed separately. Nevertheless, the present work would be very helpful for understanding the flexibility of short DNAs and the controversy on the high flexibility of short DNAs. The obtained empirical formula of apparent persistence length might be practically useful.

# ACKNOWLEDGEMENT

We are grateful to Profs. Shi-Jie Chen, Haiping Fang and Wenbing Zhang for valuable

discussions, and Chang Shu for facility assistance. We are also grateful for Prof. Xian-Wu Zou and Ya-Zhou Shi for critical reading the manuscript. This work was supported by the National Key Scientific Program (973)-Nanoscience and Nanotechnology (No. 2011CB933600), the National Science Foundation of China grants (11074191, 11175132 and 11374234), the Program for New Century Excellent Talents (Grant No. NCET 08-0408). One of us (Y. Y. Wu) also thanks financial supports from the interdisciplinary and postgraduate programs under the Fundamental Research Funds for the Central Universities.

# Figure Captions

**Figure 1** (a) An illustration for the end-to-end distance $R_{ee}$ and the contour length $L$ of the 50-bp atomistic DNA helix; see also Table 1 and Fig. S1. (b) The instantaneous end-to-end distance $R_{ee}$ (red and green lines) *versus* MD simulation time from two independent MD simulations for the 50-bp DNA helix. (c) The averaged value of end-to-end distance over every $\Delta t$ by $\int_{t-\Delta t/2}^{t+\Delta t/2} R_{ee}(t')dt'/\Delta t$ (red and green lines) *versus* MD simulation time from two independent MD simulations for the 50-bp DNA helix. Here, $\Delta t$ =500ps. The central line denotes the mean value $\overline{R}_{ee}$ of end-to-end distance in equilibrium (MD time is larger than 100ns for the 50-bp DNA). The analyses on the DNA conformations were performed with the program Curves+.[59]

**Figure 2** (a) The normalized probability distributions of end-to-end distance $R_{ee}$ (red) and contour length $L$ (blue) distribution curves for the 5-bp, 10-bp, 15-bp, 20-bp, 25-bp, 30-bp, 40-bp, and 50-bp DNA helices, respectively (from left to right). Note that the distributions are not perfectly normal and the deviation from normal distribution is slight, as discussed in main text. (b) The mean end-to-end distance $R_{ee}$ (red) and contour length $L$ (blue) as functions of the DNA length (bps). The mean values of $R_{ee}$ and $L$ are simply calculated by averaging over all the possible conformations in equilibrium. The slopes of end-to-end distance $R_{ee}$ and contour length $L$ are ~3.22Å and ~3.30 Å per base pair, respectively.

**Figure 3** The variances $\sigma_L^2$ of contour length $L$ (a, b) and the variances $\sigma_R^2$ of end-to-end distance $R_{ee}$ (c) as functions of DNA length (bps). Solid lines: all-atomistic MD; dash lines: MC simulation with the worm-like chain model where the stretching modulus $k$ and persistence length $l_p$ are shown in the figures. $\sigma_L^2 = \overline{(L-\overline{L})^2}$ and $\sigma_R^2 = \overline{(R_{ee}-\overline{R}_{ee})^2}$ were calculated by averaging the values over all the conformations in equilibrium. It is noted that the WLC model does indeed reproduce the desired $l_p$ with a high degree of accuracy, as shown in supplementary material.[41]

**Figure 4** (a) The distribution of rise per base pair along the 20-bp, 30-bp and 50-bp DNAs (from top to bottom). (b) The variance distribution of rise per base pair along the 20-bp, 30-bp and 50-bp DNAs (from top to bottom). For panels (a) and (b), the mean rise $\overline{r}_i$ and rise variance $\sigma_r^2(i) = \overline{(r_i-\overline{r}_i)^2}$ along the short DNAs were calculated by averaging the values over all the conformations in equilibrium. (c) The mean rise per base pair for short DNAs as a function of DNA length (bps). (d) The mean variance of rise per base pair for short DNAs as a function of DNA length (bps). It is noted that in panels (a) and (b), the lines were fitted with quadratic function rather than linear horizontal function since the former can give better fits with smaller deviation for the points. For the 50-bp DNA, the fitting deviations with quadratic function are 0.11432 and 0.01818 for the rise per base pair and the rise variance, while the fitting deviations with the linear horizontal function are 0.12907 and 0.020695, respectively. The distribution of rise per base pair and variances (averaged over 6 bps) along the 20-bp, 30-bp, and 50-bp DNAs were also shown in Fig. S7 of supplementary materials.[41]

**Figure 5** (a) The distribution of bending angle (over 6 bps) along the 20-bp, 30-bp, and 50-bp DNAs (from top to bottom). (b) The distribution of bending angle variance (over 6 bps) along the 20-bp, 30-bp, and 50-bp DNAs (from top to bottom). For panels (a) and (b), the bending angle $\bar{\theta}_i$ and rise variance $\sigma_\theta^2(i) = \overline{(\theta_i - \bar{\theta}_i)^2}$ along the short DNAs were calculated by averaging the values over all the conformations in equilibrium. (c) The mean bending angle (over 6 bps) for short DNAs as a function of DNA length (bps). (d) The mean variance of bending angle (over 6 bps) for short DNAs as a function of DNA length (bps).

**Figure 6** (a) The sharp bending (≥30º over 6 bps) probability along the 30-bp and 50-bp DNAs from all-atom MD simulations (blue symbols) and MC simulations with the WLC model with the same apparent persistence length from the MD simulations (red lines; $l_p$=45nm and 38nm for the 50-bp and 30-bp DNAs, respectively). It is noted that in panel (a), the blue lines were fitted with quadratic function rather than linear horizontal function since the former can give better fits with smaller deviation for the points. For the 50-bp DNA, the fitting deviation with quadratic function is 0.00471391 while the fitting deviations with the linear horizontal function is 0.00558103; (b) The mean sharp bending (≥30º over 6 bps) probability for short DNA helices as a function of DNA length (bps). The sharp bending probability over 6 bps for a short DNA was calculated with averaging the probability in each 6-bp section over the whole length of the DNAs. (c) A snapshot of the 40-bp DNA to show the sharp bending sites denoted by the circles. (d) The base-pair rise distribution along the 40-bp DNA helix for the conformation shown in panel (c). The central axis of the DNA is determined with the program Curves+.[59]

**Figure 7** (a) An illustration for the region of high ion charge density (larger than 0.02 $e$/Å$^3$) around a 40-bp DNA helix with a sharp bending. (b) The distribution of binding ion charge per bp in the cylindrical cell with radius of 12Å around the bent DNA helix. (c) The distribution of binding ion charge in concave side in the cylindrical cell with radius of 12Å around the bent DNA helix. (d) The distribution of binding ion charge in convex side in the cylindrical cell with radius of 12Å around the bent DNA helix.

**Figure 8** (a) The apparent persistence lengths $l_p$ of short DNAs as a function of DNA length (bps). The apparent persistence length $l_p$ was calculated from all-atomistic MD with Eq. 3 (●) and Eq. 4 (◆), respectively. The experimental apparent persistence length $l_p$ (■) was calculated from the experimental data of the radius of gyration of the short DNAs by (26) $\langle R_g^2 \rangle = L l_p / 3 - l_p^2 + 2 l_p^3 / L - 2 l_p^4 (1 - \exp(-L / l_p)) / L^2$, and $R_g^2$ is corrected by involving DNA radius $R_{\text{DNA}}$:[26] $R_{g,DNA}^2 = R_g^2 + R_{DNA}^2/2$, where the radius of DNA $R_{DNA}$=1.1nm.[26] The dash line is the persistence length calculated from the WLC model with $l_p$=45nm. The solid line is the empirical formula of Eq. 5. (b) The variance $\sigma_R^2$ of end-to-end distance $R_{ee}$ of short DNAs as a function of DNA length (bps). Solid line: from all-atomistic MD; Dash line: from the WLC model with $k$=1320pN and a length-dependent $l_p$ of Eq. 5.

**Figure 9** (a) The variance $\sigma_R^2$ of end-to-end distance $R_{ee}$ as a function of DNA length for isolated short DNAs (full symbols) and the "inner" short DNAs with the same length which are taken from the middle of the 50-bp DNA (open symbols), respectively. (b) The variance $\sigma_R^2$ of $R_{ee}$ as a function of DNA length for the short DNAs with excluding 2, 4 and 6 bps at each helix end. Here, DNA length indicates that of the short DNAs after 2, 4, and 6 bps at each end were removed. (c) The apparent persistence length of the short DNAs with excluding 2, 4, 6, 8 and 10 base pairs at each helix end. Here, DNA length indicates the original length of the short DNAs before 2, 4, 6, 8 and 10 bps at each end were removed.

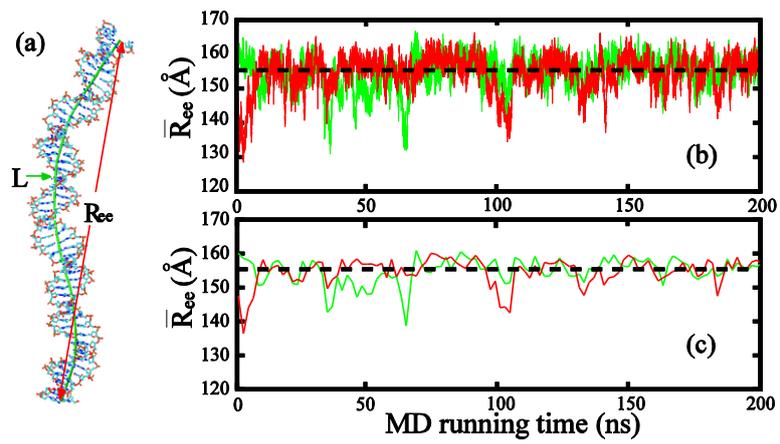

**Figure 1**

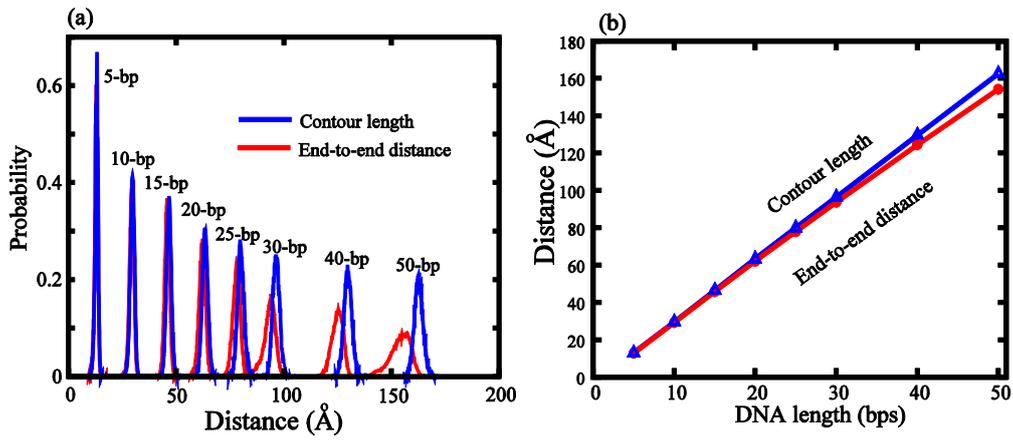

**Figure 2**

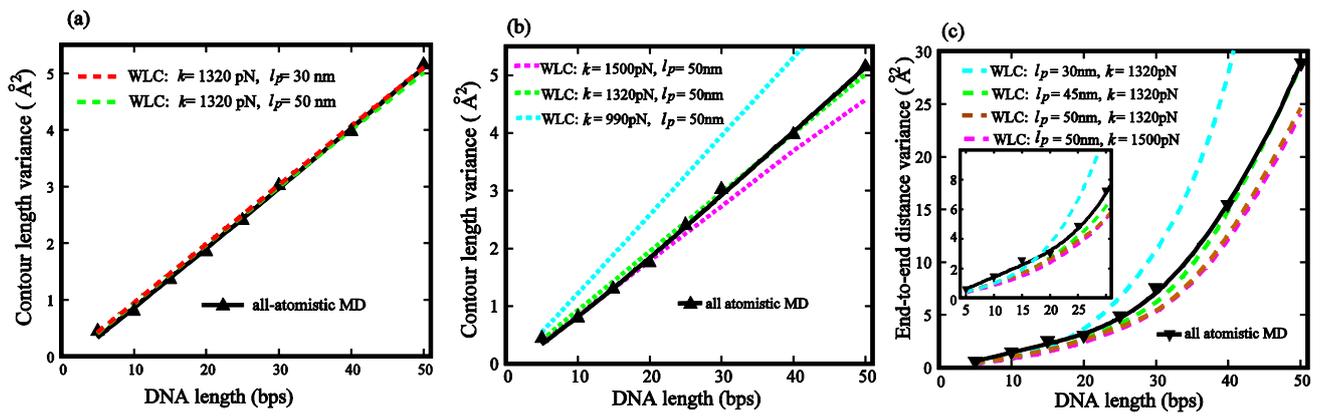

**Figure 3**

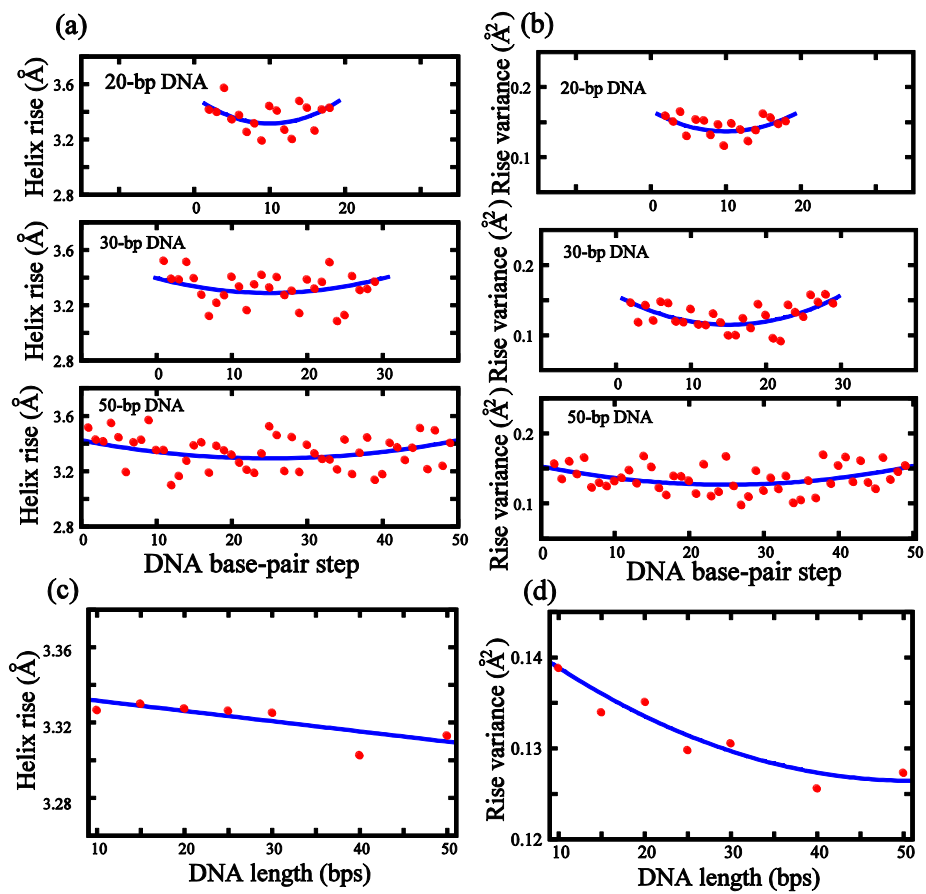

**Figure 4**

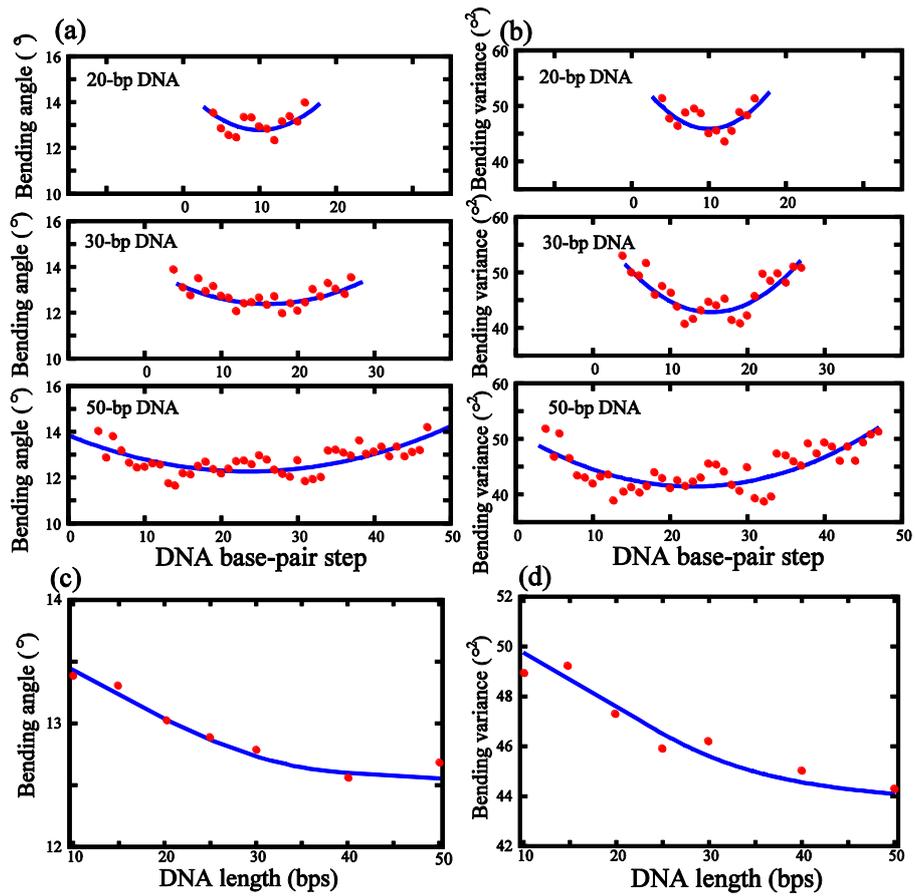

**Figure 5**

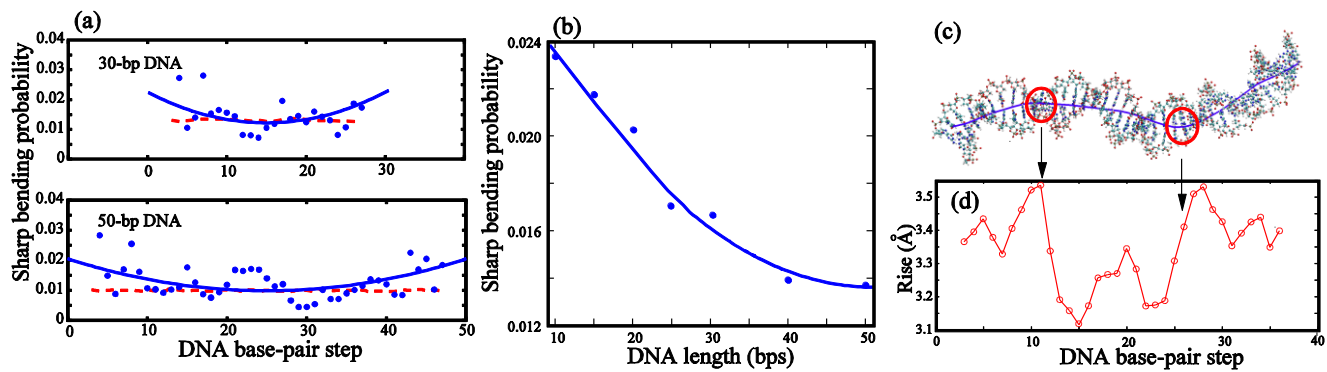

**Figure 6**

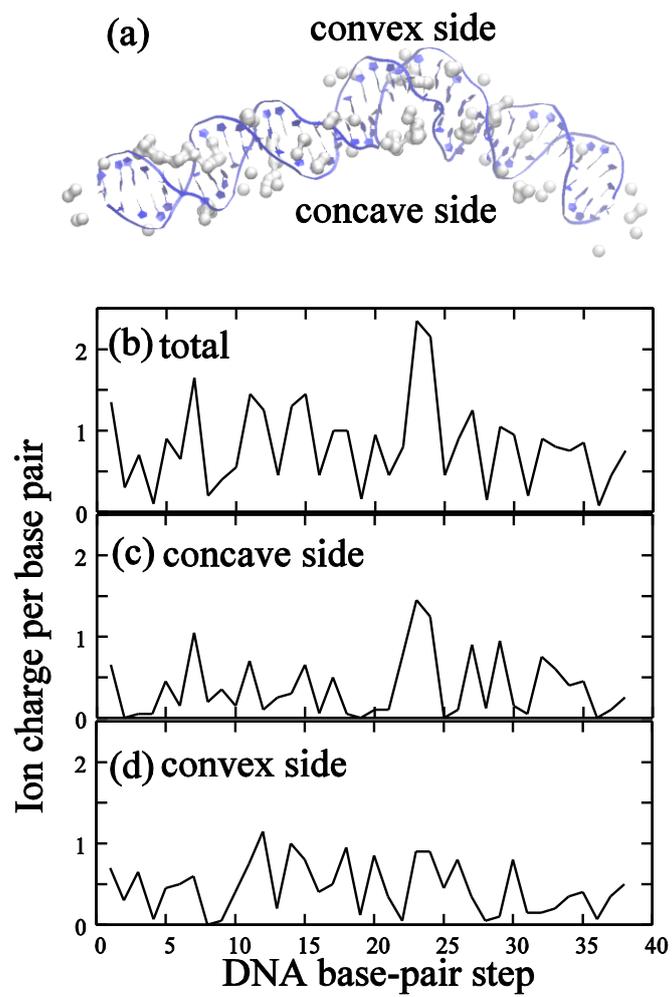

**Figure 7**

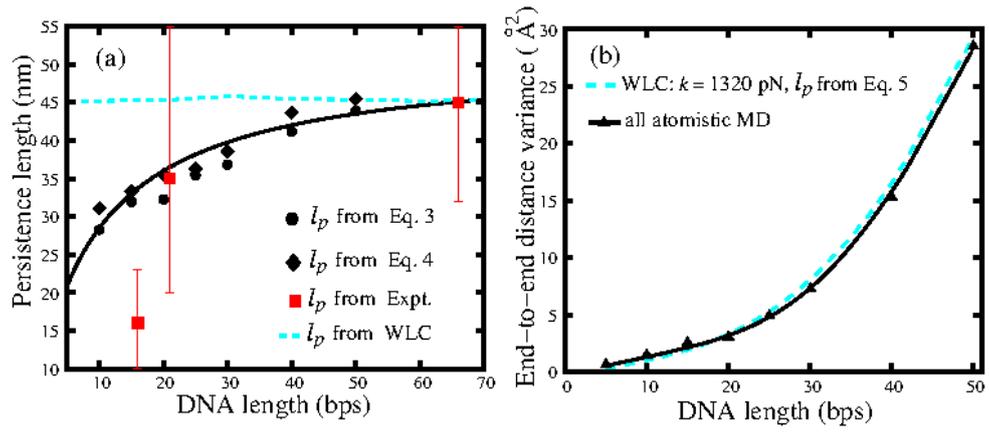

**Figure 8**

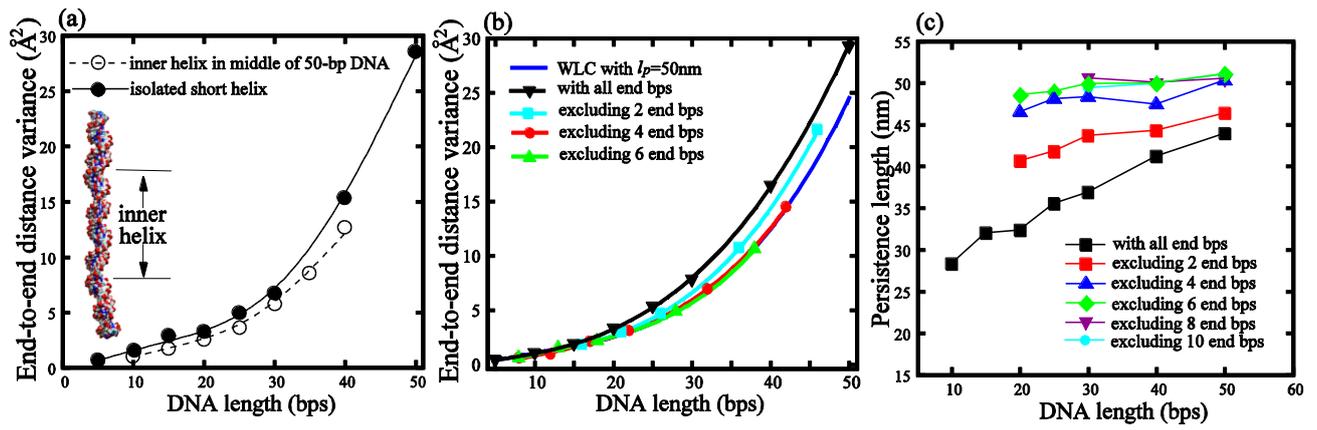

**Figure 9**

**Table 1**  DNA sequences used in the study.[a]

| helix length (bps) | Sequences | Simulation box size ($L_x \times L_y \times L_z$) Å³ | $C_S$ (M)[b] | Predicted $T_m$ (ºC)[c] |
|---|---|---|---|---|
| 5 | 5'-GCAGC-3'<br>CGTCG | 72×72×74 | 0.00866 | 50.5 |
| 10 | 5'-GCATCTGGGC-3'<br>CGTAGACCCG | 72×72×74 | 0.00866 | 77.8 |
| 15 | 5'-CGACTCTACGGAAGG-3'<br>GCTGAGATGCCTTCC | 72×72×74 | 0.00866 | 81.2 |
| 20 | 5'-CGACTCTACGGCATCTGCGC-3'<br>GCTGAGATGCCGTAGACGCG | 80×80×90 | 0.00577 | 90.1 |
| 25 | 5'-CGACTCTACGGAAGGGCATCTGCGC-3'<br>GCTGAGATGCCTTCCCGTAGACGCG | 85×86×102 | 0.00446 | 92.8 |
| 30 | 5'-CGACTCTACGCAAGGTCTCGGACTACGCGC-3'<br>GCTGAGATGCCTTCCAGAGCCTGATGCGCG | 91×91×132 | 0.00304 | 93.6 |
| 40 | 5'-CGACTCTACGGAAGGGCATCCTTCGGGCATCACTACGCGC-3'<br>GCTGAGATGCCTTCCCGTAGGAAGCCCGTAGTGATGCGCG | 91×91×167 | 0.00240 | 96.8 |
| 50 | 5'-CGACTCGACTCTACGGAAGGGCATCCTTCGGGCATCACTACGCGCCGCGC-3'<br>GCTGAGCTGAGATGCCTTCCCGTAGGAAGCCCGTAGTGATGCGCGGCGCG | 91×91×192 | 0.00209 | 100.2 |

[a]The sequences of short DNAs are selected according to the recent SAXS experimental sequences[27] to yield normal B-form DNA helices, and contain all the dinucleotide base pairs;[44-46]

[b]$C_S$ is the strand concentration (in M) which was calculated according to the corresponding simulation box.

[c]The melting temperatures were estimated from the nearest neighbor model based on the measured thermodynamic parameters.[44-46]